\pdfoutput=1
\documentclass[12pt,preprint]{aastex}
\usepackage{graphicx}
\usepackage{natbib}
\usepackage{aas_macros}
\usepackage[section] {placeins}
\usepackage{subfigure}
\usepackage{float}
\usepackage{color}

\def\msun{{\rm\,M_\odot}}

\def\SBunit{{\rm erg\, s^{-1} cm^{-2} arcsec^{-2}}}

\def\La{{\rm L_{Ly\alpha}}}
\def\msun{{\rm\,M_\odot}}

\newcommand{\kms}{\, {\rm km\, s}^{-1}}

\newcommand{\mpc}{\, {\rm Mpc}}
\newcommand{\kpc}{\, {\rm kpc}}
\newcommand{\hmpc}{\, h^{-1} \mpc}

\newcommand{\lya}{Ly$\alpha$ }

\newcommand{\hi}{\mbox{H\,{\scriptsize I}\ }}
\newcommand{\heii}{\mbox{He\,{\scriptsize II}\ }}

\def\h2{${\rm\,H_2}$}

\makeatletter 

\def\hi{\noindent \hangindent=2.5em}

\def\kpc{{\rm\,kpc}}

\def\mpc{{\rm\,Mpc}}

\def\kms{{\rm\,km/s}}
\def\msun{{\rm\,M_\odot}}

\def\vol#1  {{{#1}{\rm,}\ }}
\def\lya{{\rm Ly}\alpha}

\def\hi{{H\thinspace I}\ }

\def\heii{{He\thinspace II}\ }

\newcount\refno
\newcount\rfno
\def\eq{$^{\the\refno\ }$\advance\refno by 1}
\def\ad{\advance\rfno by 1}

\def\clock{\count0=\time \divide\count0 by 60
     \count1=\count0 \multiply\count1 by -60 \advance\count1 by \time
     \number\count0:\ifnum\count1<10{0\number\count1}\else\number\count1\fi}

\def\myputfigure#1#2#3#4#5%
{\hskip0.03\textwidth\vskip#5pt
\makebox[0pt]{\hskip#2in
\includegraphics[width=#3\textwidth]{#1}}\vskip#4pt\hfill}

\newcount\refno
\newcount\rfno
\def\eq{$^{\the\refno\ }$\advance\refno by 1}
\def\ad{\advance\rfno by 1}

\definecolor{burntorange}{rgb}{1,0.4,0.2}

\begin{document}

\title{Nature of Lyman Alpha Blobs: Powered by Extreme Starbursts}
 
\author{
Renyue Cen$^{1}$ and Zheng Zheng$^2$
} 
 
\footnotetext[1]{Princeton University Observatory, Princeton, NJ 08544;
 cen@astro.princeton.edu}
\footnotetext[2]{University of Utah, Department of Physics and Astronomy, Salt Lake City, UT 84112; zhengzheng@astro.utah.edu}

\begin{abstract} 
We present a new model for the observed Ly$\alpha$ blobs (LABs) 
within the context of the standard cold dark matter model.
In this model, LABs are the most massive halos with the strongest clustering (proto-clusters) 
undergoing extreme starbursts in the high-z universe.
Aided by calculations of detailed radiative transfer of $\lya$ photons through ultra-high resolution ($159$pc) large-scale ($\ge 30$Mpc) 
adaptive mesh-refinement cosmological hydrodynamic simulations with galaxy formation,
this model is shown to be able to, for the first time, reproduce simultaneously the global 
$\lya$ luminosity function and luminosity-size relation of the observed LABs. 
Physically, a combination of dust attenuation of $\lya$ photons within galaxies,
clustering of galaxies, and complex propagation of $\lya$ photons through
circumgalactic and intergalactic medium gives rise to 
the large sizes and frequently irregular isophotal shapes of LABs that are observed. 
A generic and unique prediction of this model is that there should be  
strong far-infrared (FIR) sources within each LAB,
with the most luminous FIR source likely representing the gravitational center of the proto-cluster,
not necessarily the apparent center of the $\lya$ emission of the LAB or the most luminous optical source.
Upcoming ALMA observations should unambiguously test this prediction.  
If verified, LABs will provide very valuable laboratories for studying formation
of galaxies in the most overdense regions of the universe at a time 
when global star formation is most vigorous.

\end{abstract}


\section{Introduction}

The physical origin of spatially extended (tens to hundreds of kiloparsecs)
luminous ($\La\ge 10^{43}$erg/s) Ly$\alpha$ sources, also known as Ly$\alpha$ blobs (LABs)
first discovered more than a decade ago \citep[e.g.,][]{1996Francis, 1999Fynbo,1999Keel,2000Steidel}, 
remains a mystery.
By now several tens of LABs have been found 
\citep[e.g.,][]{2004Matsuda, 2005Dey, 2006Saito, 2009Smith, 2011Matsuda}.
One fact that has confused the matter considerably 
is that they appear to be associated with a very diverse galaxy population,
including regular Lyman break galaxies (LBGs)
\citep[e.g.,][]{2004Matsuda},
ultra-luminous infrared galaxies (ULIRGs) 
and sub-millimeter galaxies (SMGs) \citep[e.g.,][]{2001Chapman, 2005Geach, 2007Geach, 2007Matsuda, 2011Yang},
unobscured \citep[e.g.,][]{2003Bunker, 2004Weidinger}
and obscured quasars \citep[e.g.,][]{2004BasuZych, 2007Geach, 2009Smith},
or either starbursts or obscured quasars 
\citep[e.g.,][]{2009Geach, 2009Scarlata,2011Colbert}.
An overarching feature, however, is that
the vast majority of them are associated with massive halos or rich large-scale
structures that reside in dense parts of the Universe and will likely evolve to become
rich clusters of galaxies by $z=0$
\citep[e.g.,][]{2000Steidel, 2004Chapman, 2004Matsuda,2004Palunas, 2006Matsuda,2008Prescott, 
2009Matsuda, 2009Yang, 2009Webb, 2010Weijmans, 2011Matsuda, 2011Erb, 2011bYang, 2011Zafar}.
Another unifying feature is that LABs are strong infrared emitters. 
For instance, most of the 35 LABs with size $>30$ kpc identified by 
\citet[][]{2004Matsuda} in the SSA 22 region have been detected in deep Spitzer
observations \citep[][]{2009Webb}.

Many physical models of LABs have been proposed.
A leading contender is the gravitational cooling radiation model
in which gas that collapses inside a host dark matter halo releases a significant 
fraction of its gravitational binding energy in Ly$\alpha$ line emission 
\citep[e.g.,][]{2000Haiman, 2001Fardal, 2003Birnboim, 2006Dijkstra, 2006Yang, 2009Dijkstra, 2010Goerdt,2010FG,2012Rosdahl}.
The strongest observational support for this model comes from 
two LABs that appear not to be associated with any
strong AGN/galaxy sources \citep[][]{2006Nilsson, 2008Smith},
although lack of sub-mm data in the case of \citet[][]{2006Nilsson}
and a loose constraint of $\le 550\msun$yr$^{-1}$ ($3\sigma$) in the case of
\citet[][]{2008Smith} both leave room to accommodate AGN/galaxies powered models.
Another tentative support is claimed to come from 
the apparent positive correlation between velocity width (represented by the 
full width at half maximum, or FWHM, of the line) and 
Ly$\alpha$ luminosity \citep[][]{2008Saito},
although the observed correlation FWHM $\propto \La$ appears to be much steeper than
expected (approximately) FWHM $\propto\La^{1/3}$ for virialized systems.
Other models include photoionization of cold dense, spatially extended gas by obscured quasars 
\citep[e.g.,][]{2001Haiman, 2009Geach}, by population III stars \citep[e.g.,][]{2006Jimenez},
or by spatially extended inverse Compton X-ray emission \citep[e.g.,][]{2003Scharf},
emission from dense, cold superwind shells 
\citep[e.g.,][]{2000Taniguchi, 2003Ohyama, 2004Mori, 2005Wilman, 2007Matsuda},
or a combination of photoionization and gravitational cooling radiation \citep[e.g.,][]{2005Furlanetto}.

The aim of this writing is, as a first step, to explore a simple star formation based model in 
sufficient details to access its physical plausibility and self-consistency,
through detailed $\lya$ radiative transfer calculations utilizing 
a large set of massive ($\ge 10^{12}\msun$) starbursting galaxies 
from an ultra-high resolution ($\sim 110 h^{-1}$pc), cosmological, adaptive mesh refinement (AMR) hydrodynamic
simulation at $z=3.1$.
The most critical, basically the only major, free parameter in our model 
is the magnitude of dust attenuation.
Adopting the observationally motivated trend that higher SFR galaxies have higher dust attenuation,
with an overall normalization that seems plausible (e.g., we assume that $\sim 5$\% of $\lya$ photons
escape a galaxy of ${\rm SFR}=100\msun$~yr$^{-1}$),
the model can successfully reproduce the global $\lya$ luminosity function and the luminosity-size relation of LABs.
To our knowledge this is the first model that is able to achieve this.
The precise dependence of dust attenuation on SFR is not critical, within a reasonable range,
and hence the results are robust.

In this model we show that LABs at high redshift correspond to proto-clusters containing the most massive galaxies/halos
in the universe.
Within each proto-cluster, all member galaxies contribute collectively 
to the overall $\lya$ emission, giving rise to the diverse geometries 
of the apparent contiguous large-area LAB emission,
which is further enhanced by projection effects due to other galaxies that 
are not necessarily in strong gravitational interactions with the main galaxy (or galaxies),
given the strong clustering environment of massive halos in a hierarchical universe.
This prediction that LABs should correspond to the most overdense regions in 
the universe at high redshift is fully consistent with the observed 
universal association of LABs with high density peaks 
(see references above). 
The relative contribution to the overall $\lya$ emission from each individual galaxy
depends on a number of variables, including dust attenuation of $\lya$ photons within the galaxy
and propagation and diffusion processes through its complex circumgalactic medium and the intergalactic medium.
Another major predictions of this model is that 
a large fraction of the stellar (and AGN) optical and ultraviolet (UV) radiation 
(including Ly$\alpha$ photons)
is reprocessed by dust and emerges as infrared (IR) radiation,
consistent with observations of ubiquitous strong infrared emission from LABs.
We should call this model simply ``starburst model" (SBM), encompassing those with or without contribution from central AGNs. 
This model automatically includes emission contribution from gravitational cooling radiation, which
is found to be significant but sub-dominant compared to stellar radiation.
Interestingly, we also find that $\lya$ emission originating from nebular emission (rather than the stellar emission), 
which includes contribution from gravitational binding energy due to halo collapse,
is more centrally concentrated than that from stars.

One potentially very important prediction is that in this model 
the Ly$\alpha$ emission from photons that escape to us
is expected to contain significant polarization signals. 
Although polarization radiative transfer calculations will be performed to detail the polarization signal in a future study,
we briefly elaborate the essential physics and latest observational advances here.
One may broadly file all the proposed models into two classes in terms of the spatial distribution of 
the underlying energy source: central powering or in situ.
Starburst galaxy and AGN powered models
belong to the former, whereas gravitational cooling radiation model belongs to the latter.
A smoking gun test between these two classes of models is  
the polarization signal of the Ly$\alpha$ emission.
In the case of a central powering source (not necessarily a point source)
the Ly$\alpha$ photons diffuse out, spatially and in frequency, through optically thick medium and escape by 
a very large number of local resonant scatterings in the Ly$\alpha$ line profile core and
a relatively smaller number of scatterings in the damping wings with long flights.
Upon each scattering a Ly$\alpha$ photon
changes its direction, location and frequency, dependent upon the geometry, density and 
kinematics of the scattering neutral hydrogen atoms.
In idealized models with central powering significant linear polarizations of tens of percent 
on scales of tens to hundreds of kiloparsecs are predicted  
and the polarization signal strength increases with radius
\citep[e.g., ][]{1998Lee, 1999Rybicki, 2008Dijkstra}. 
On the other hand, in situ radiation from the gravitational cooling model
is not expected to have significant polarizations (although detailed modeling will be needed
to quantify this) or any systematic radial trend,
because thermalized cooling gas from (likely) filaments will emit
Ly$\alpha$ photons that are either not scattered significantly or have no preferential orientation or 
impact angle with respect to the scattering medium.

An earlier attempt to measure polarization of LABd05 at $z=2.656$ 
produced a null detection \citep[][]{2011cPrescott}. 
A more recent observation by \citet[][]{2011Hayes}, for the first time,
detected a strong polarization signal tangentially oriented (almost forming a complete ring) 
from LAB1 at $z=3.05$, whose strength increases with radius from the LAB center,
a signature that is expected from central powering;
they found the polarized fraction (P) of 20 percent at a radius of 45 kpc. 
\citet[][]{2011Hayes} convincingly demonstrate their detection and,
at the same time, explain the consistency of their result 
with the non-detection by \citet[][]{2011cPrescott}, 
if the emission from LABd05 is in fact polarized, 
thanks to a significant improvement in sensitivity and spatial resolution
in \citet[][]{2011Hayes}.
This latest discovery lends great support to models with central powering, including SBM,
independent of other observational constraints that may or may not differentiate 
between the two classes of models or between models in each class. 
But we stress that detailed polarization calculations will be needed to enable 
statistical comparisons.


The outline of this paper is as follows.
In \S 2.1 we detail simulation parameters and hydrodynamics code,
followed by a description of our Ly$\alpha$ radiative transfer method in \S 2.2.
Results are presented in \S 3
with conclusions given in \S 4.

\section{Simulations}\label{sec: sims}

\subsection{Hydrocode and Simulation Parameters}

We perform cosmological simulations with the AMR 
Eulerian hydro code, Enzo 
\citep[][]{1999bBryan, 2009Joung}.  
First we ran a low resolution simulation with a periodic box of $120~h^{-1}$Mpc 
(comoving) on a side.
We identified a region centered on a cluster of mass of $\sim 3\times 10^{14}\msun$ at $z=0$.
We then resimulate with high resolution of the chosen region embedded
in the outer $120h^{-1}$Mpc box to properly take into account large-scale tidal field
and appropriate boundary conditions at the surface of the refined region.
The refined region 
has a comoving size of $21\times 24\times 20h^{-3}$Mpc$^3$ 
and represents $1.8\sigma$ matter density fluctuation on that volume.
The dark matter particle mass in the refined region is $1.3\times 10^7h^{-1}\msun$.
The refined region is surrounded by three layers (each of $\sim 1h^{-1}$Mpc) of buffer zones with 
particle masses successively larger by a factor of $8$ for each layer, 
which then connects with
the outer root grid that has a dark matter particle mass $8^4$ times that in the refined region.
We choose the mesh refinement criterion such that the resolution is 
always better than $111h^{-1}$pc (physical), corresponding to a maximum mesh refinement level of $13$ at $z=0$.
The simulations include
a metagalactic UV background
\citep[][]{1996Haardt},  
and a model for shielding of UV radiation by neutral hydrogen 
\citep[][]{2005Cen}.
They include metallicity-dependent radiative cooling \citep[][]{1995Cen}.
Our simulations also solve relevant gas chemistry
chains for molecular hydrogen formation \citep[][]{1997Abel},
molecular formation on dust grains \citep[][]{2009Joung},
and metal cooling extended down to $10~$K \citep[][]{1972Dalgarno}.
Star particles are created in cells that satisfy a set of criteria for 
star formation proposed by \citet[][]{1992CenOstriker}.
Each star particle is tagged with its initial mass, creation time, and metallicity; 
star particles typically have masses of $\sim$$10^6\msun$.

Supernova feedback from star formation is modeled following \citet[][]{2005Cen}.
Feedback energy and ejected metal-enriched mass are distributed into 
27 local gas cells centered at the star particle in question, 
weighted by the specific volume of each cell, which is to mimic the physical process of supernova
blastwave propagation that tends to channel energy, momentum and mass into the least dense regions
(with the least resistance and cooling).
We allow the entire feedback processes to be hydrodynamically coupled to surroundings
and subject to relevant physical processes, such as cooling and heating. 
The total amount of explosion kinetic energy from Type II supernovae
for an amount of star formed $M_{*}$
with a Chabrier initial mass function (IMF) is $e_{SN} M_* c^2$ (where $c$ is the speed of light)
with  $e_{SN}=6.6\times 10^{-6}$.
Taking into account the contribution of prompt Type I supernovae,
we use $e_{SN}=1\times 10^{-5}$ in our simulations.
Observations of local starburst galaxies indicate
that nearly all of the star formation produced kinetic energy 
is used to power galactic superwinds \citep[e.g.,][]{2001Heckman}. 
Supernova feedback is important primarily for regulating star formation
and for transporting energy and metals into the intergalactic medium.
The extremely inhomogeneous metal enrichment process
demands that both metals and energy (and momentum) are correctly modeled so that they
are transported in a physically sound (albeit still approximate 
at the current resolution) way.
The kinematic properties traced by unsaturated metal lines in damped Lyman-alpha 
systems (DLAs) are
extremely tough tests of the model, which is shown to agree well with observations \citep[][]{2012Cen}.

We use the following cosmological parameters that are consistent with 
the WMAP7-normalized \citep[][]{2010Komatsu} $\Lambda$CDM model:
$\Omega_M=0.28$, $\Omega_b=0.046$, $\Omega_{\Lambda}=0.72$, $\sigma_8=0.82$,
$H_0=100 h \,{\rm km\, s}^{-1} {\rm Mpc}^{-1} = 70 \,{\rm km\, s}^{-1} {\rm Mpc}^{-1}$ and $n=0.96$.
This simulation has been used \citep[][]{2011cCen} to  
quantify partitioning of stellar light into optical and infrared light, through
ray tracing of continuum photons in a dusty medium that is based on self-consistently
computed metallicity and gas density distributions.

We identify galaxies in our high resolution simulations using the HOP algorithm 
\citep[][]{1999Eisenstein}, operated on the stellar particles, which is tested to be robust
and insensitive to specific choices of concerned parameters within reasonable ranges.
Satellites within a galaxy are clearly identified separately.
The luminosity of each stellar particle at each of the Sloan Digital Sky Survey (SDSS) five bands 
is computed using the GISSEL stellar synthesis code \citep[][]{Bruzual03}, 
by supplying the formation time, metallicity and stellar mass.
Collecting luminosity and other quantities of member stellar particles, gas cells and dark matter 
particles yields
the following physical parameters for each galaxy:
position, velocity, total mass, stellar mass, gas mass, 
mean formation time, 
mean stellar metallicity, mean gas metallicity,
star formation rate,
luminosities in five SDSS bands (and various colors) and others.
At a spatial resolution of $159$pc (physical) with nearly 5000 well resolved galaxies at $z\sim 3$,
this simulated galaxy catalog presents an excellent (by far, the best available) 
tool to study galaxy formation and evolution.

\subsection{$\lya$ Radiative Transfer Calculation}

The AMR simulation resolution is $159$pc at $z=3$.
For each galaxy we produce a cylinder of size 
$(2R_{\rm vir})\times (2R_{\rm vir}) \times (42R_{\rm vir})$ on a uniform grid 
of cell size $318$pc, where $R_{\rm vir}$ is the virial radius of the host halo.
The purpose of using the elongated geometry is to incorporate the line-of-sight 
structures. Subsequently, in our $\lya$ radiative transfer calculation, the 
line-of-sight direction is set to be along the longest dimension of the cylinder.
In each cell of a cylinder $\lya$ photon emissivities are computed,
separately from star formation and cooling radiation.
The luminosity of $\lya$ produced by star formation is computed
as $\La = 10^{42}[{\rm SFR}/(\msun {\rm yr}^{-1})]\, {\rm erg\, s^{-1}}$ 
\citep[][]{2005Furlanetto}, where SFR is the star formation rate in the cell. 
The $\lya$ emission from cooling radiation is computed with the gas
properties in the cell by following the rates of excitation and ionization.

With $\lya$ emissivity, neutral hydrogen density, temperature, and 
velocity in the simulations, a Monte Carlo code \citep[][]{ZM02} is 
adopted to follow the $\lya$ radiative transfer. The code has been
recently used to study $\lya$ emitting galaxies \citep[][]{Zheng10,Zheng11a,
Zheng11b}. 
In our radiative transfer calculation, the number of $\lya$ photons drawn 
from a cell is proportional to the total $\lya$ luminosity in the cell, with
a minimum number of 1000, and each photon is given a weight in order to 
reproduce the luminosity of the cell. $\lya$ photons associated with star 
formation and cooling radiation are tracked separately so that we can 
study their final spatial distributions. 
For each photon, the scattering with neutral hydrogen atoms and the subsequent 
changes in frequency, direction, and position are followed until it escapes 
from the simulation cylinder. More details about the code can be found in 
\citet[][]{ZM02} and \citet[][]{Zheng10}.

The pixel size of the $\lya$ images from the radiative transfer calculation is chosen 
to be equal to $318$pc, corresponding to $0.04\arcsec$.
We smooth the $\lya$ images with 2D Gaussian kernels
to match the resolutions in \citet[][]{2011Matsuda} for detecting and
characterizing LABs from observation. In \citet[][]{2011Matsuda},
the area of an LAB is the isophotal area with a threshold surface 
brightness $1.4\times 10^{-18}\SBunit$ in the narrowband image smoothed to 
an effective seeing of FWHM 1.4$\arcsec$ 
(slightly different from \citealt{2004Matsuda}, where FWHM=1$\arcsec$), 
while the $\lya$ 
luminosity is computed with the isophotal aperture in the FWHM=1$\arcsec$ 
image. We define LABs in our model by applying a friends-of-friends 
algorithm to link the pixels above the threshold surface brightness in
the computed $\lya$ images, with the area and luminosity computed from smoothed
images with FWHM=1.4$\arcsec$ and FWHM=1$\arcsec$, respectively.

\section{Results}

The SBM model that we study here in great detail
may appear at odds with available observations at first sight.
In particular, the LABs often lack close correspondence 
with galaxies in the overlapping fields and their centers
are often displaced from the brightest galaxies in the fields.
As we show below, these puzzling features are in fact 
exactly what are expected in the SBM model.
The reasons are primarily three-fold.
First, LABs universally arise in large halos with a significant number
of galaxies clustered around them.
Second, dust attenuation renders the amount of Ly$\alpha$ emission emerging
from a galaxy dependent substantially sub-linearly on star formation rate.
Third, the observed Ly$\alpha$ emission, in both amount and three-dimensional (3D) 
location,
originating from each galaxy depends on complex scattering processes subsequently.

\begin{figure*}[h!]
\vskip -0.0cm
\centering
\hskip 0.1cm
\resizebox{7.00in}{!}{\includegraphics[angle=0]{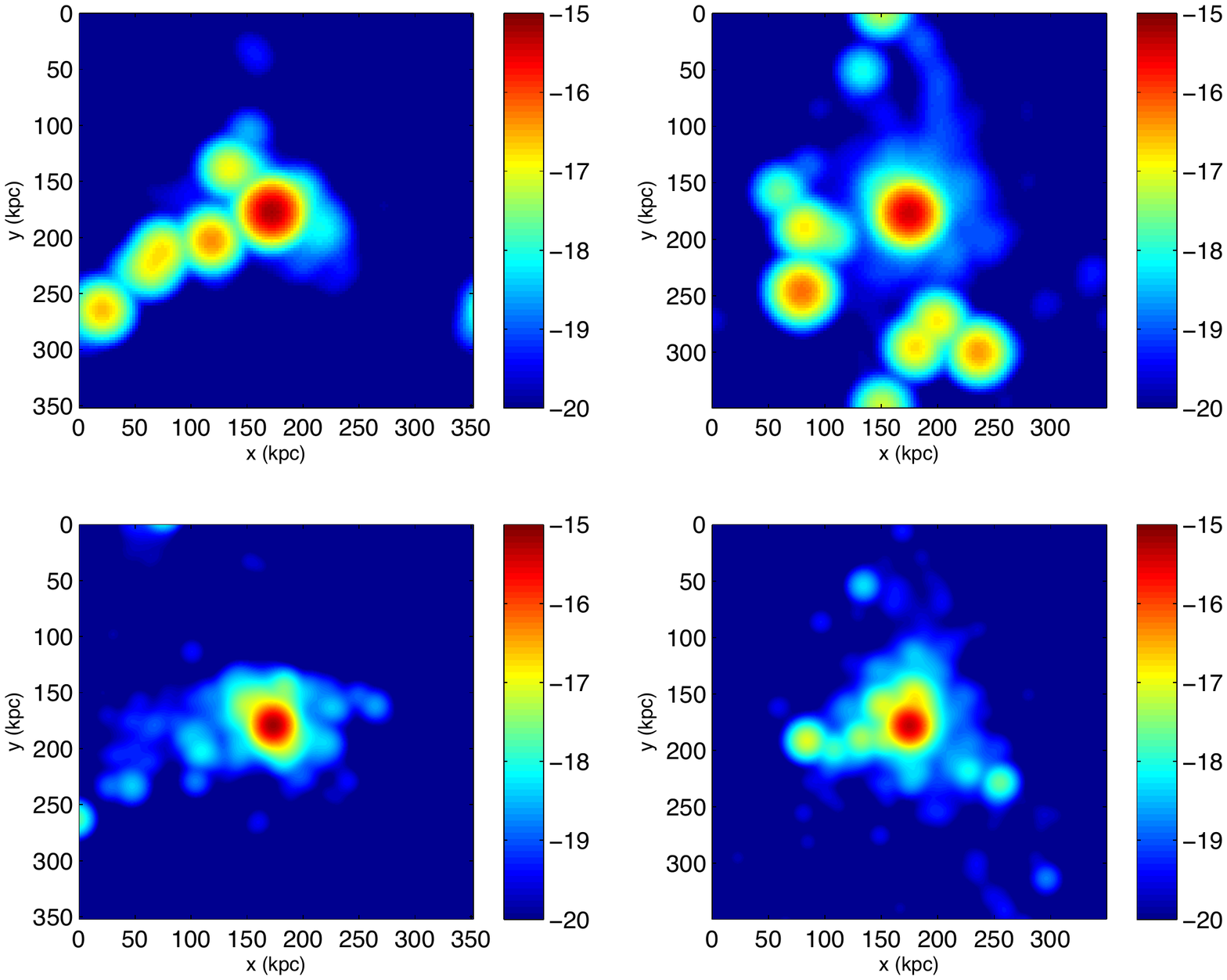}}
\vskip -0.0cm
\caption{
Two examples: left (a) and right (b) columns.
See the caption below with columns (c) and (d).
}
\label{fig:map2}
\end{figure*}

\addtocounter{figure}{-1} 

\begin{figure*}[h!]
\vskip -0.0cm
\centering
\hskip 0.1cm
\resizebox{7.00in}{!}{\includegraphics[angle=0]{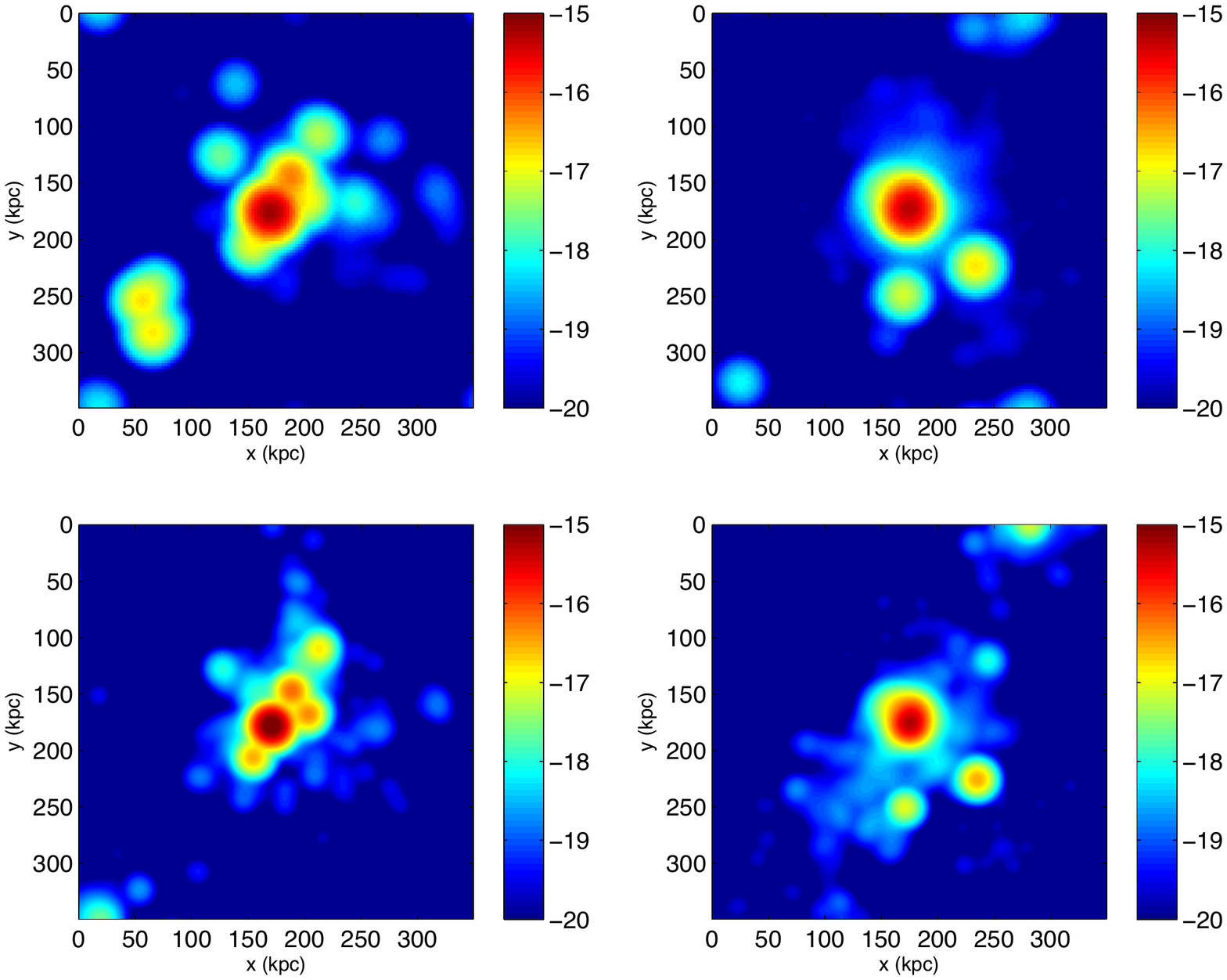}}
\vskip -0.0cm
\caption{
Two more examples: left (c) and right (d) columns.
The four columns (a,b,c,d) show the logarithm of Ly$\alpha$ surface brightness maps (in units of $\SBunit$)
for four randomly selected large galaxies of virial masses both
exceeding $10^{12}\msun$ at $z=3.1$ with the primary galaxy 
centered on their respective panel.
For each column the bottom panel is obtained, if one only includes 
galaxies within $\pm R_{\rm vir}$ of the primary galaxy along the line of sight, where $R_{\rm vir}$ is the 
virial radius of the primary galaxy.
The top panel is obtained, including all galaxies within $\pm 10h^{-1}$Mpc comoving of the primary galaxy 
along the line of sight.  The length shown is in physical kpc.
The effects of dust and faint sources have not
been included yet in these plots (see the text for more details).
}
\label{fig:map2}
\end{figure*}

\if(0) {
\begin{figure*}[ht]
\plottwo{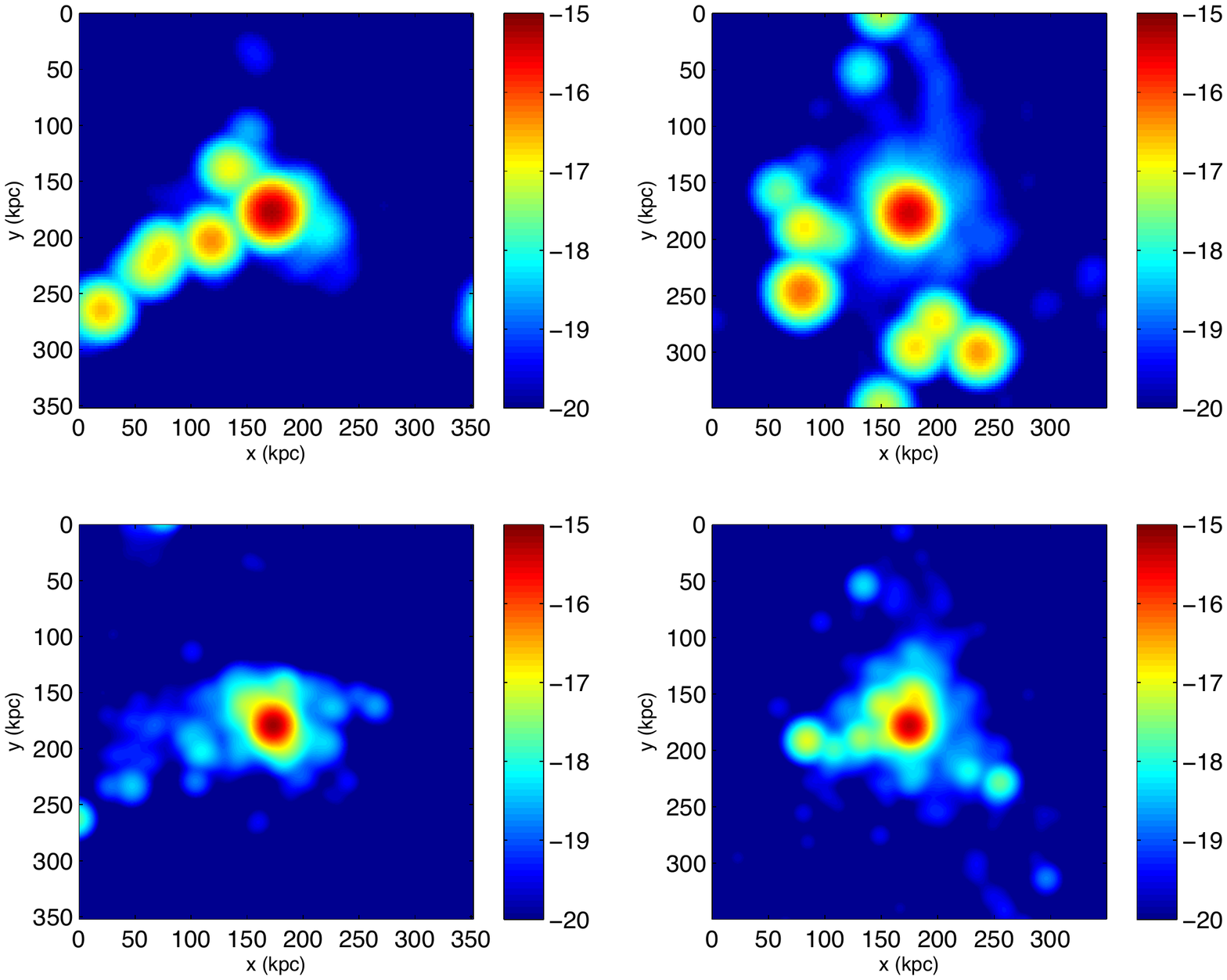}{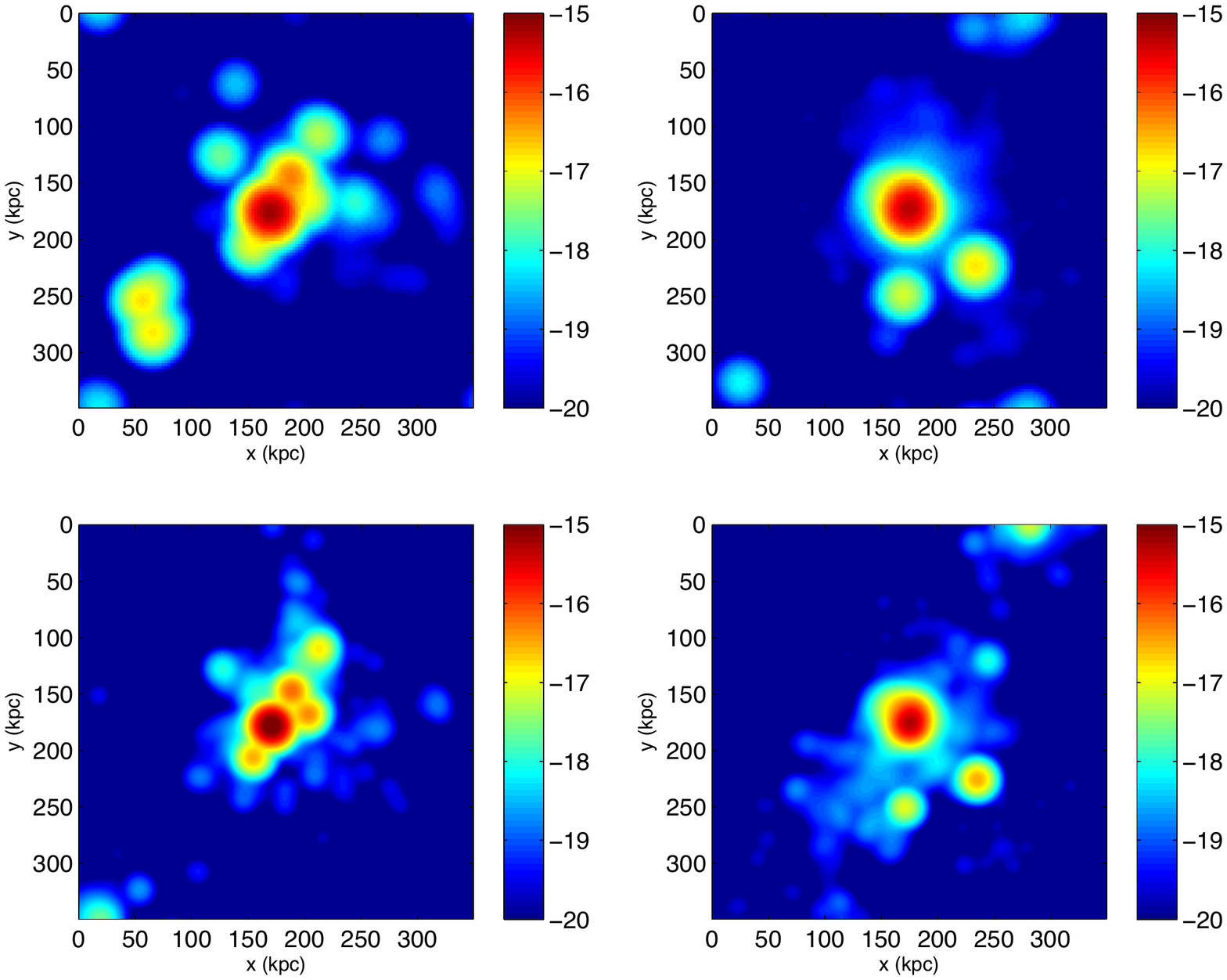}
\caption{
\label{fig:map2}
The four columns show the logarithm of Ly$\alpha$ surface brightness maps
(in units of $\SBunit$)
for four randomly selected large galaxies of virial masses both
exceeding $10^{12}\msun$ at $z=3.1$ with the primary galaxy
centered on their respective panel.
For each column the bottom panel is obtained, if one only includes
galaxies within $\pm R_{\rm vir}$ of the primary galaxy along the line of sight, where $R_{\rm vir}$ is the
virial radius of the host halo of the primary galaxy.
The bottom panel is obtained, including all galaxies within $\pm 10h^{-1}$Mpc 
(comoving) of the primary galaxy
along the line of sight.
The length shown is in physical kpc.
The galaxies are taken from the simulation with resolution of
$110h^{-1}$pc \citet[][]{2011eCen}. The effects of dust and faint sources have not
been included yet in these plots (see the text for more details). 
{\color{blue} 
\bf ZZ: should make the x- and y-labels, and axis tick labels larger -- they are 
hard to see.}
}
\end{figure*}
}
\fi

\subsection{Effects Caused by Galaxy Clustering}
We find that large-scale structure and clustering of galaxies
play a fundamental role in shaping all aspects of LABs,
including two-dimensional line-of-sight velocity structure,
line profile and $\lya$ image in the sky plane.
To illustrate this, Figure \ref{fig:map2} shows 
Ly$\alpha$ surface brightness maps (after the radiative transfer calculation)
for four randomly selected galaxies with virial mass of the central galaxy 
exceeding $10^{12}\msun$ at $z=3.1$.
We find that Ly$\alpha$ emission stemming from stellar radiation
dominate over the gas cooling by about 10:1 to 4:1 in all relevant cases.
We also find that the Ly$\alpha$ emission due to gas cooling is at least
as centrally concentrated as from the stellar emission for each galaxy.
From this figure it has become clear that
large-scale structure and projection effects are 
instrumental to rendering the appearance of LABs in all aspects (image as well as spectrum).
One could see that, for example in the top-left panel of Figure \ref{fig:map2},
the approximately linear structure aligned in the direction of lower-left to upper-right
is composed of three additional galaxies 
that are well outside the virial radius of the primary galaxy but from projected structures.
At the $1.4\times 10^{-18} \SBunit$ detection isophotal 
contours of \citet[][]{2004Matsuda} and \citet[][]{2011Matsuda} for LABs,
the entire linear structure may be identified as a single LAB.
This rather random example is strikingly reminiscent of the observed 
LAB structures \citep[e.g.,][]{2009Matsuda,2011Erb,2011bYang}.
Interestingly, depending on which galaxy is brighter and located on the front or back,
the overall Ly$\alpha$ emission of the LAB may show a variety of line profiles.
For example, it could easily account for a broad/brighter blue side in the line profile,
as noted by \citet[][]{2006Saito} for some of the observed LABs,
which was originally taken as supportive evidence for the gravitational cooling radiation model.
Furthermore, it is not difficult to envision that 
the overall velocity width of an LAB does not necessarily reflect 
the virial velocity of a virialized system and may display a wide range
from small (masked by caustics effect) to large (caused by either large virial 
velocities, infall velocities, or Hubble expansion). 
A detailed spectral analysis will be presented elsewhere.

For the results shown in Figure \ref{fig:map2} 
we have not included dust effect, contributions from small galaxies ($M_h < 10^{9.5}\msun$) 
that are not properly captured in our simulation due to finite resolution,
and instrumental noise.
We now describe how we include these important effects.

\subsection{Taking into Account Faint, Under-resolved Sources}

\label{sec:faintsource}

Although the resolution of our simulations is high, it is still finite and 
small sources are incomplete. We find that the star formation rate (SFR) function
in the simulation flattens out at 3$\msun\, {\rm yr}^{-1}$ toward lower SFR at $z=2-3$ \citep[][]{2011cCen},
which likely means that sources with SFR$<3 \msun\, {\rm yr}^{-1}$ are unresolved/under-resolved and hence incomplete in the simulations. 
Since these low SFR sources that cluster around large galaxies 
contribute to the $\lya$ emission of LABs, it is necessary to include them in our modeling.
For this purpose, we need to sample their SFR distribution and spatial distribution inside halos.

First, we need to model the luminosity or SFR distribution of the faint, unresolved 
sources.
In each LAB-hosting halo in the simulations, the number of (satellite) sources 
with SFR$>$3$\msun\, {\rm yr}^{-1}$ is found to be proportional to the halo 
mass $M_h$. 
Observationally, the faint end slope 
$\alpha$ of the UV luminosity function of star forming galaxies is 
$\sim -1.8$ \citep[e.g.,][]{2009Reddy}.
Given this faint end slope, the contribution due to faint, unresolved sources
is weakly convergent.  
As a result, the overall contribution from faint sources do not strongly depend on 
the faint limit of the correction procedure.
We find that the conditional SFR function $\phi(L;M_h)$ of faint sources 
(SFR$<3\msun\, {\rm yr}^{-1}$) in halos can be modeled as
\begin{equation}
\label{eqn:CSFRF}
\phi(L;M_h) = \frac{dN(M_h)}{dL} 
          = \frac{-(\alpha+1)}{L_{\rm th}}
            \left(\frac{L}{L_{\rm th}}\right)^\alpha \frac{M_h}{M_1}, 
\end{equation}
where $L$ represents the SFR and $L_{\rm th}=3\msun\, {\rm yr}^{-1}$,
$\alpha=-1.8$, and $M_1=10^{12}\msun$. This conditional SFR function
allows us to draw SFR for faint sources to be added in our model.

We now turn to the spatial distribution of faint sources.
In the simulation the spatial distribution (projected to the sky plane) 
of satellite sources in halos is found to closely follow a power-law with 
a slope of $-2$. This is in good agreement with the observed small scale 
slope of the projected two-point correlation function of LBGs 
\citep[][]{2005Ouchi}. 
There is some direct observational evidence that there are faint UV sources distributed within the LAB radii. 
\citet{2012Matsuda} perform stacking analysis
of $z\sim 3.1$ $\lya$ emitters and protocluster LBGs, 
showing diffuse $\lya$ profile in the stacked $\lya$ image. Interestingly, the 
profiles in the stacked UV images appear to be extended to scales of tens of 
kpc (physical) for the most luminous $\lya$ sources or for sources in 
protoclusters, suggesting contributions from faint, starforming galaxies. 

We add the contribution from faint sources to  
post-processed unsmoothed $\lya$ images from radiative transfer modeling as follows. 
For each model LAB, we draw the number
and SFRs of faint sources in the range of 0.01--3$\msun {\rm yr}^{-1}$ 
based on the conditional SFR distribution in equation~(\ref{eqn:CSFRF}). Then
we distribute them in the unsmoothed $\lya$ image in a radial range of 
0.01--1$R_{\rm vir}$ by following the power-law distribution with slope
$-2$. The faint sources can be either added as point or extended sources
in $\lya$ emission. If added as point sources, they would be smoothed with 
a 2D Gaussian kernel of FWHM=1.4$\arcsec$ or 1$\arcsec$ when defining LAB size 
and luminosity. In our fiducial model, each faint source is added 
as an exponential disk with scale length of 3$\arcsec$ to approximate the 
radiative transfer effect, which is consistent with the observed diffuse 
emission profile of star-forming galaxies \citep[][]{2011Steidel}. 
We find that our final conclusion does not sensitively depend on our choice of the faint source $\lya$ profile.

\begin{figure*}[ht]
\plotone{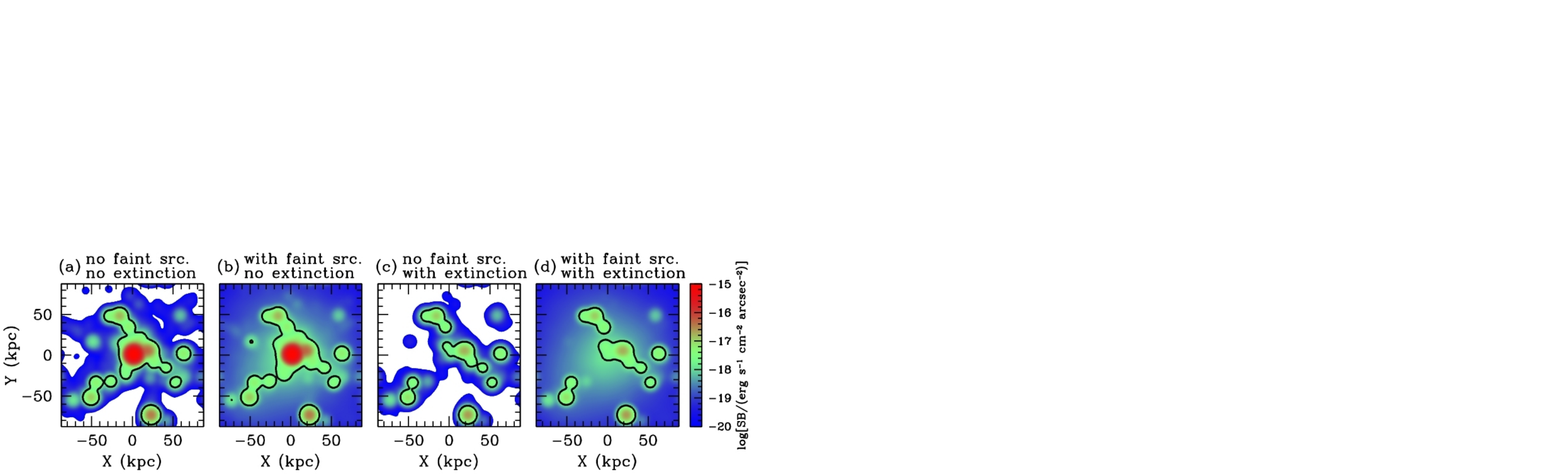}
\caption{
\label{fig:isophot}
An LAB under different model assumptions. The model LAB shown in this 
example resides in the most massive host halo in our simulation 
($\sim 5\times 10^{12}\msun$) at $z=3.1$. The $\lya$ images are smoothed to correspond to
seeing of FWHM=1.4$\arcsec$. 
In each panel, the black contour is the 
isophotal level of $1.4\times 10^{-18} \SBunit $, the surface brightness 
threshold used in observation to define LABs
\citep[][]{2004Matsuda,2011Matsuda}. Panels (a)--(d) enumerate the combinations
of adding faint sources and extinction. Panel (a) is the initial case 
without faint sources and without extinction. Panel (d) corresponds to
the case with faint sources added and with extinction considered, which we
regard as the favored model. See the text for more details.
}
\end{figure*}

In Figure~\ref{fig:isophot}, panel (a) shows the surface brightness and
the $1.4\times 10^{-18}\SBunit$ isphotal contour for a model LAB without 
including the faint sources, while panel (b) is the case with faint sources. 
We see that the size of the LAB defined by the isophotal aperture does not 
change much. 
If the $\lya$ 
emission of each faint sources is more concentrated, e.g., close to a point 
source in the unsmoothed image, the LAB size can increase a little bit. 
Therefore, in both panels (a) and (b), the size is mainly determined by
the central bright source. 
However, as will be described in the next subsection,
including the effect of dust extinction will suppress the contribution of the 
central source and relatively boost that of the faint sources in determining the LAB size.

\subsection{Dust Effect}

In the cases shown in Figure \ref{fig:map2},
the central galaxies each have SFR that exceeds $100\msun$~yr$^{-1}$ and
is expected to be observed as a luminous infrared galaxy (LIRG) or ULIRG \citep[][]{1996Sanders}. 
This suggests that dust effects are important and have to be taken into account.

In general, there are two types of effects of dust on $\lya$ emission from
star-forming galaxies. The 
first one is related to the production of $\lya$ phtons. 
Dust attenuates ionizing photons in star-forming galaxies. Since $\lya$ 
photons come from reprocessed ionizing photons, the attenuation by dust leads
to a lower $\lya$ luminosity in the first place. Second, after being produced,
$\lya$ photons can be absorbed by dust during propagation. 
A detailed 
investigation needs to account for both effects self-consistently, and we
reserve that for a future study.  

In \citet[][]{2011cCen} the dust obscuration/absorption is considered in a 
self-consistent way, with respect to luminosity functions observed in UV and FIR 
bands.
The modelling uses detailed ray tracing with dust obscuration model
based on that of our own Galaxy \citep[][]{2011Draine} and
extinction curve taken from \citet[][]{1989Cardelli}.
While the simultaneous match of both UV and FIR luminosity functions at $z=2$ without introducing
additional free parameters is an important validation of the physical realm of our simulations,
it is not necessarily directly extendable to the radiative transfer of $\lya$ photons.
Nevertheless, it is reasonable to adopt a simple optical depth approach as follows for our present purpose,
normalized by relevant observations,
as follows.

For each galaxy we suppress the initial intrinsic Ly$\alpha$ emission, 
by applying a mapping $\La$ to $\La \exp{[-\tau({\rm SFR})]}$, where
the ``effective" optical depth $\tau({\rm SFR})$ is intended to account for extinction of $\lya$ photons
as a function of SFR. 
We stress that this method is approximate and its validation 
is only reflected by the goodness of our model fitting the observed properties of LABs. 
We adopt $\tau({\rm SFR}) = 0.2 [{\rm SFR}/(\msun{\rm yr^{-1}})]^{0.6}$.
In reality, in addition, it may be that there is a substantial scatter in $\tau({\rm SFR})$ at a fixed SFR.
We ignore such complexities in this treatment.
The adopted trend that higher SFR galaxies have larger optical depths is fully consistent 
with observations \citep[e.g.,][]{2009bNilsson}. 
At intrinsic ${\rm SFR}=100 \msun{\rm yr}^{-1}$ the escaped $\La$ luminosity is equivalent to 
${\rm SFR}=5 \msun {\rm yr}^{-1}$, whereas 
at intrinsic ${\rm SFR}=10 \msun {\rm yr}^{-1}$ the escaped $\La$ luminosity is equivalent to ${\rm SFR}=4.5 \msun {\rm yr}^{-1}$.
It is evident that the scaling of the emerging $\La$ luminosity on intrinsic SFR 
is substantially weakened with dust attenuation.
In fact, it may be common that, due to dust effect, the optical luminosity of a galaxy 
does not necessarily positively correlate with its intrinsic SFR, or the most luminous source
in $\lya$ does not necessarily correspond to the highest SFR galaxy within an LAB.
As a result, a variety of image appearance and mis-matches between the LAB centers
and the most luminous galaxies detected in other bands may result,
seemingly consistent with the anecdotal observational evidence mentioned 
in the introduction.

The effect of dust on the surface brightness distribution for a model LAB
is shown in panel (c) of Figure~\ref{fig:isophot}. Compared to panel (a),
which is the model without dust effect, we see that surface brightness of
the central source is substantially reduced and the isophotal area for 
the threshold $1.4\times 10^{-18}\SBunit$ also reduces. The case in panel 
(c) does not include the contribution from faint sources. 
In general, taking into account dust effect in our Ly$\alpha$ radiative transfer calculation,
the central galaxies tend to make reduced (absolutely and relative to other smaller nearby galaxies)
contributions to the Ly$\alpha$ surface brightness maps
and in fact the center of each LAB may or may not coincide with the primary galaxy
that would likely be a ULIRG in these cases, which is again reminiscent of some observed LABs.
In the next subsection,  we describe the modeling results of combining all the above effects.

\subsection{Final LABs with All Effects Included}

By accounting for the line-of-sight structures, the unresolved faint sources,
and the dust effect, we find that the observed properties of LABs
can be reasonably reproduced by our model. 

In panel (d) of Figure~\ref{fig:isophot}, we add the faint sources and apply 
the dust effect. Compared with the case in panel (c), where no faint sources
are added, the isophotal area increases. The central source has a 
substantially reduced surface brightness because of extinction. There appears
to be another source near the central source, which corresponds to a source
of lower SFR seen in panel (a) but with lower extinction than the central 
source. From Figure~\ref{fig:isophot}, we see that the overall effect is that
dust helps reduce the central surface brightness and faint sources help somewhat enlarge the isophotal area. 

\begin{figure*}[h!]
\hskip 0.0cm
\centering
\resizebox{5.0in}{!}{\includegraphics[angle=0]{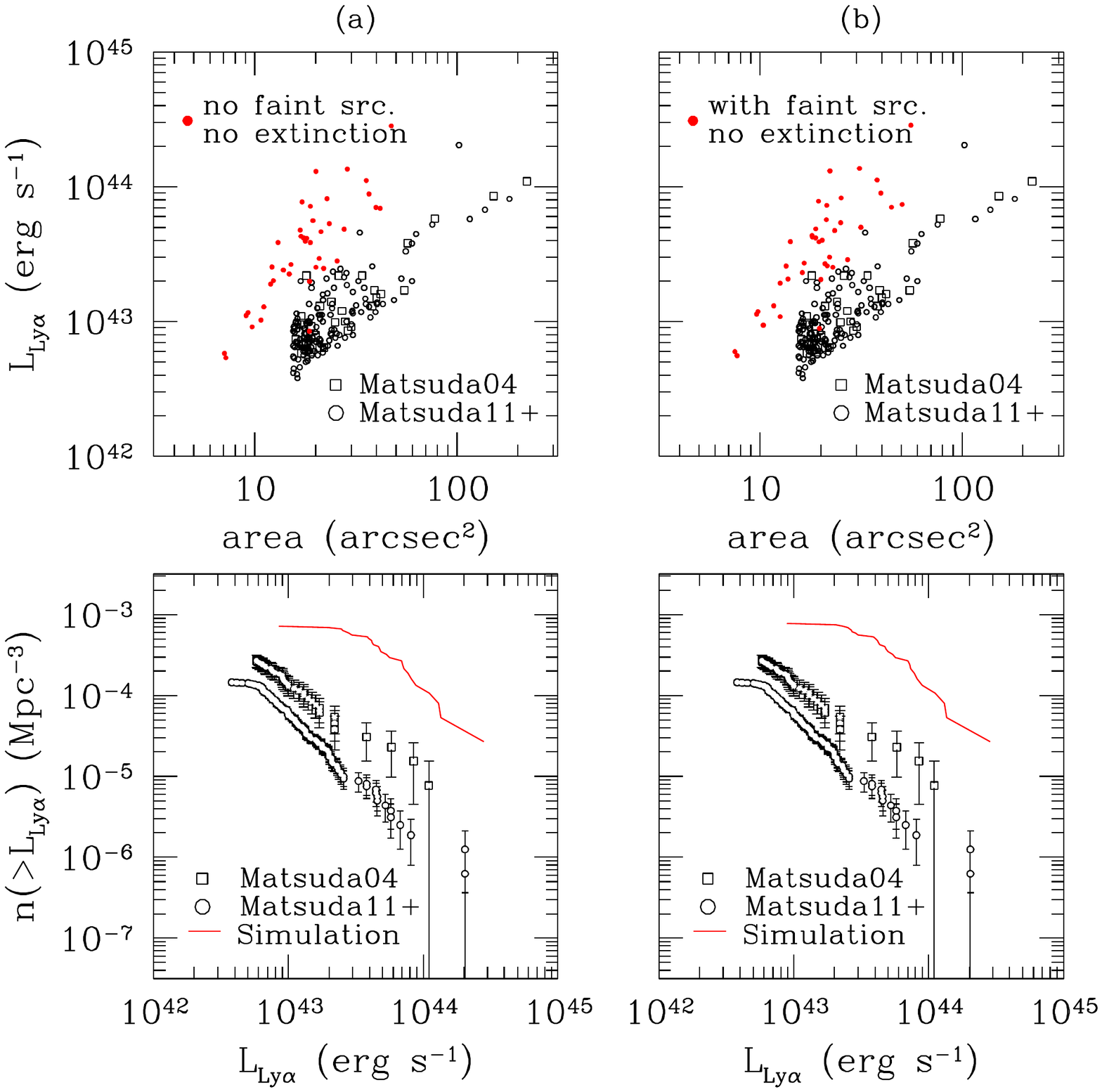}}
\vskip -3.0cm
\caption{
Model predictions under different assumptions along with observed properties of LABs.
Top panels show luminosity and size relations and bottom panels cumulative luminosity functions.
Panel (a) does not account for dust effect and contributions from faint galaxies under-resolved in our simulation.
Panel (b) includes under-resolved sources.
Observations are taken from \citet[][]{2004Matsuda} (open squares) and
\citet[][]{2011Matsuda} supplemented with new unpublished data (open circles).
Model predictions are shown as red points (top panels) and curves (bottom panels).
}
\label{fig:pilot1}
\end{figure*}
\addtocounter{figure}{-1} 
\newpage

\begin{figure*}[h!]
\hskip 0.0cm
\centering
\resizebox{5.0in}{!}{\includegraphics[angle=0]{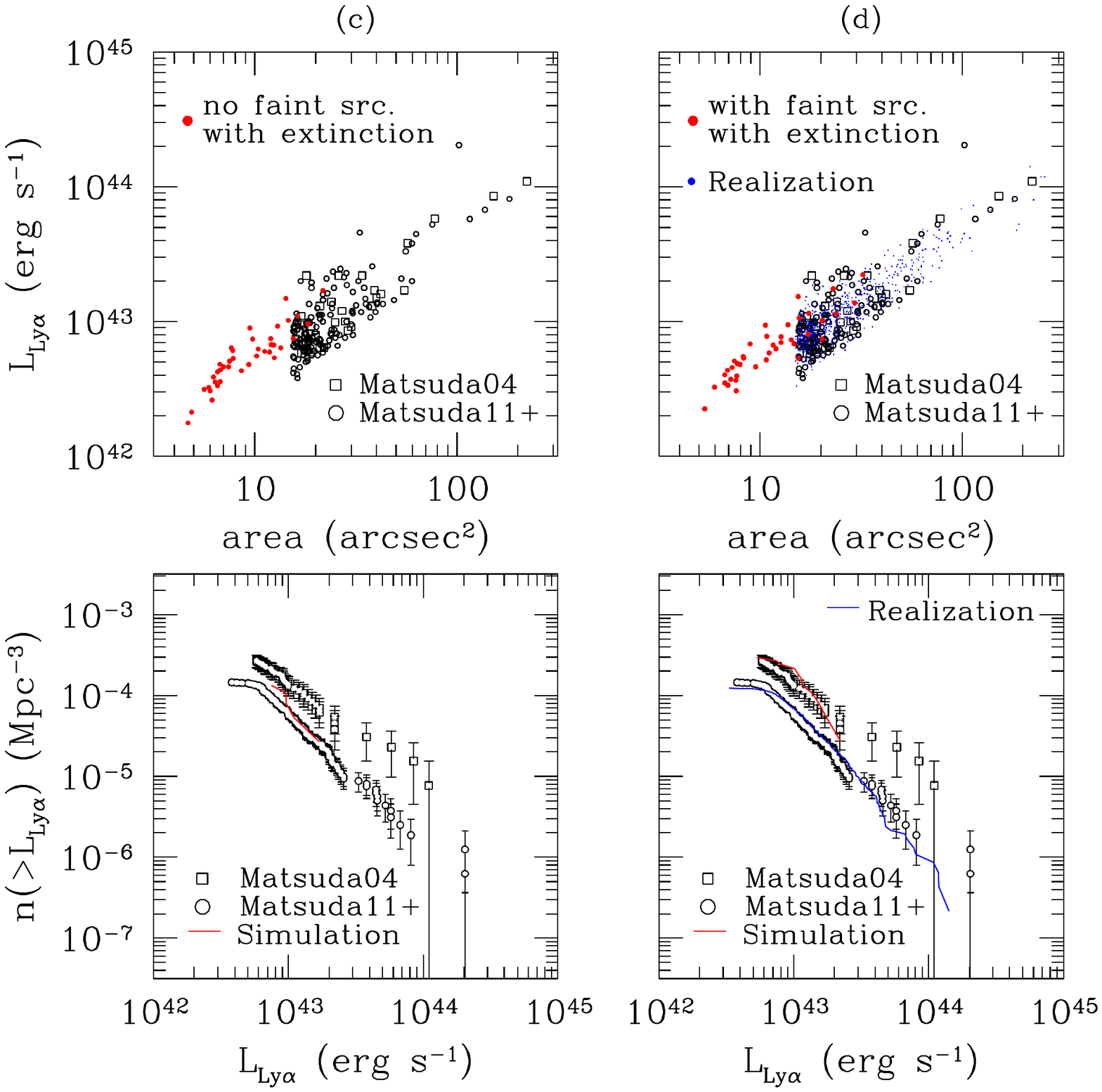}}
\vskip -3.0cm
\caption{
Continued. 
Panel (c) only includes the dust extinction effect. 
Panel (d) includes both the dust extinction and the faint sources. 
The blue dots and blue curve in panel (d) is our realization of the global LF by using the $L_{\lya}-M_h$ relation from our model and the analytic halo mass function.
}
\label{fig:pilot2}
\end{figure*}

To test the model and see the effect of different assumptions on extinction
and faint sources, we compare the model predictions with observational 
properties of LABs, shown in Figure~\ref{fig:pilot2}. In the top panels,
we compare the luminosity-size relation defined by the isophot with surface
brightness $1.4\times 10^{-18}\SBunit$. The observed data points are taken
from \citet{2004Matsuda} (open squares) and \citet{2011Matsuda} (open circles),
which has been supplemented with new, yet unpublished data (Matsuda 2012, private communications).
Note that the isophotal area is defined with FWHM=1$\arcsec$ and 
1.4$\arcsec$ images in \citet{2004Matsuda} and \citet{2011Matsuda}, 
respectively. This may partly explain that the LAB sizes are somewhat larger 
with the \citet{2011Matsuda} data points. However, the difference is not 
substantial. Our model data points follow \citet{2011Matsuda} in defining 
the luminosity and size.

In the bottom panels of Figure~\ref{fig:pilot2}, we show the cumulative $\lya$
luminosity function or abundance of LABs. The data points from 
\citet{2004Matsuda} and \citet{2011Matsuda} (supplemented with new unpublished data; Matsuda 2012, private communications)
have a large offset ($\sim$1 dex at the luminous end) 
from each other, suggesting large sample variance. The survey volumes of
\citet{2004Matsuda} and \citet{2011Matsuda} are $1.3\times 10^5 {\rm Mpc}^3$
and $1.6\times 10^6 {\rm Mpc}^3$, respectively. For comparison, the volume 
of our parent simulation from which we choose our LAB sample is only 
$3.06\times 10^4 {\rm Mpc}^3$, much smaller than the volume probed by 
observation. 

The red points in top panel (a) of Figure~\ref{fig:pilot2} come from
our model without extinction and faint sources. Compared to the observational 
data, the model predicts more or less the correct slope in the luminosity-size
relation. However, the overall relation has an offset, which means that the 
model either overpredicts the luminosity or underpredicts the size, or both. 
From the bottom panel (a), the model greatly over-predicts the LAB abundance,
showing as a vertical shift. But it can also be interpreted as an 
overprediction of the LAB luminosity, leading to a horizontal shift, which is
more likely. 
Because the central sources are bright, adding faint sources only slightly changes 
the sizes, as shown in panel (b), which leads to little improvement in solving 
the mismatches in the luminosity-size relation and in the abundance.

Once the dust extinction effect is introduced, the situation greatly improves.
Panel (c) of Figure~\ref{fig:pilot2} shows the case with extinction but without
adding faint sources. With the extinction included, the luminosity of the 
predicted LABs drops, and at the same time, the size becomes smaller. 
Now the model points agree well with observations 
at the lower end of the range of LAB luminosity ($10^{42.6}-10^{43.3}$erg/s) and size (15-30~arcsec$^2$), 
the predicted luminosity-size relation conforms to and extends the observed one to still lower luminosity and smaller size. 
The predicted abundance is much closer to the observed one, as well.

Finally, panel (d) shows the case with both extinction and faint sources 
included. Adding faint sources helps enlarge the size of an LAB, because
faint sources extends the isophot to larger radii. The luminosity also
increases by including the contribution from faint sources. As a whole, the
model data points appear to slide over the luminosity-size relation towards
higher luminosity and larger size. 
The model luminosity-size relation, although still
at the low luminosity end, is fully overlapped with the observed relation. 
The abundance at the high-luminosity end from the
model is within the range probed by observation and shows a similar slope 
as that in \citet{2004Matsuda}. 
The agreement of the luminosity function between simulations and 
\citet{2004Matsuda} is largely fortuitous, reflecting that 
the overall bias of our simulation box over the underlying matter happens to be similar to that  
of the \citet{2004Matsuda} volume over matter, provided that the model universe
is a reasonable statistical representation of the real universe.

Limited by the simulation volume, we are not able to directly simulate the full range of
the observed luminosity and size of LABs. Our model, however, reproduces the
luminosity-size relation and abundance in the low luminosity end.
The most important ingredient in our model to achieve
such an agreement with the observation is the dust extinction, which drives 
the apparent $\lya$ luminosity down into the right range. Accounting for the contribution 
of faint, unresolved sources in the simulation also plays a role in further
enhancing the sizes and, to a less extent, the luminosities of LABs.

\begin{figure}[h!]
\hskip 0.0cm
\centering
\resizebox{5.00in}{!}{\includegraphics[angle=0]{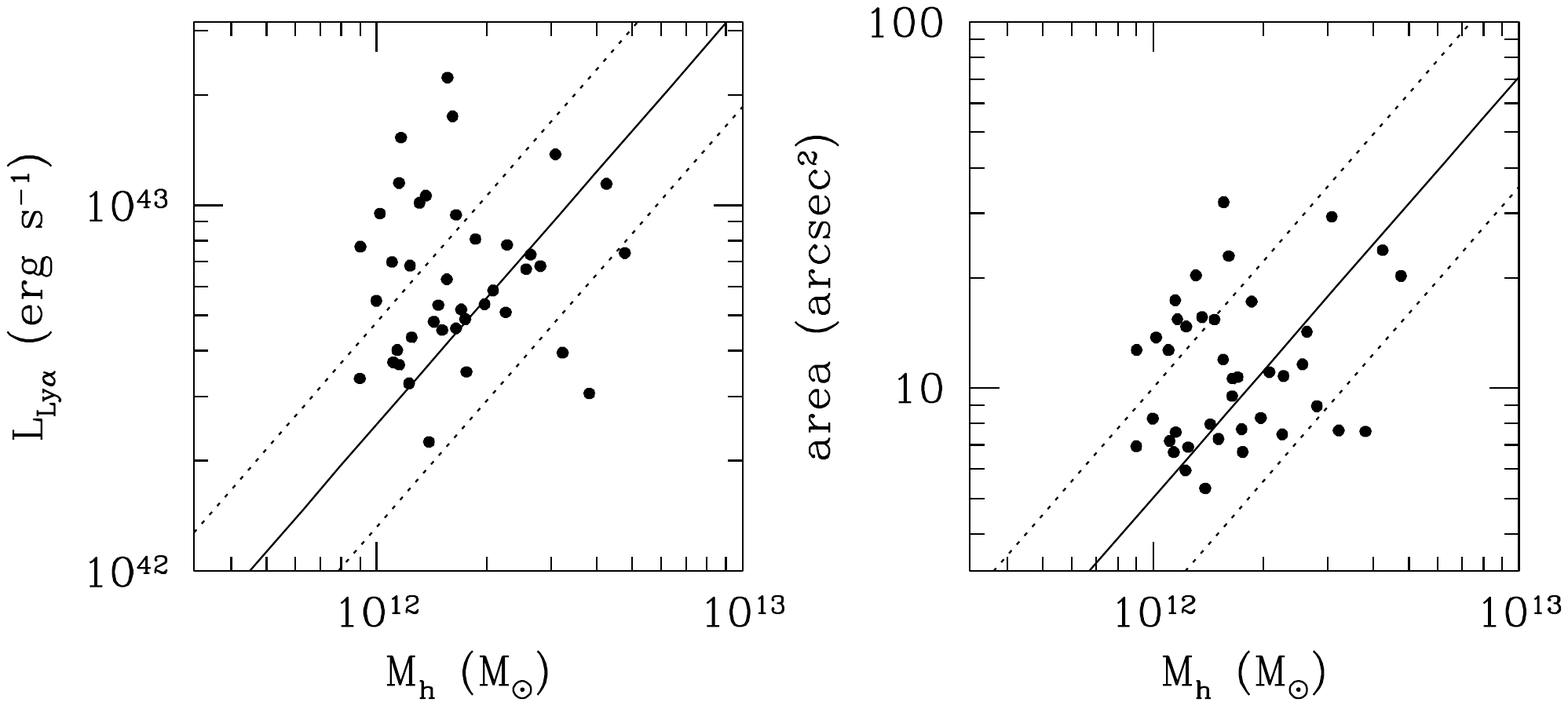}}
\vskip -9.0cm
\caption{
Dependence of LAB luminosity and size on halo mass from the model.
In each panel, the points are from our model LABs in the simulation.
The solid and dotted lines show the relation and scatter we use to 
populate halos drawn from the analytic halo mass function to compute the
expected global $\lya$ LF of LABs. See the text for more details.
}
\label{fig:halomass}
\end{figure}

To rectify the lack of high luminosity, large size LABs in our simulations due to the limited simulation volume,
we perform the following exercise.
Figure~\ref{fig:halomass} shows the $\lya$ luminosity and LAB size as a
function of halo mass from our model LABs in Figure~\ref{fig:pilot2}(d). 
Both quantities correlate with halo mass, but there is
a large scatter, which is caused by 
varying SFRs as well as different environmental effects for halos of a given mass.
The largest LABs fall into the range probed by the observational data and 
they reside in halos above $10^{12}M_\odot$.
The model suggests that the vast majority of the observed LABs should reside in proto-clusters with 
the primary halos of mass above $10^{12}M_\odot$ at $z\sim 3$ and on average larger LABs correspond to more massive halos. Note that the sources with halo mass below $10^{12}M_\odot$
is highly incomplete here.
Our results suggest an approximate relation between the halo mass
of the central galaxy and the apparent $\lya$ luminosity of the LAB:
\begin{equation}
\label{eqn:ML}
\La = 10^{42.4} \left({M_h\over 10^{12}\msun}\right)^{1.15} {\rm erg\, s^{-1}},
\end{equation}
which is shown as the solid curve in the left panel of Figure~\ref{fig:halomass}.
This relation should provide a self-consistency test of our model,
when accurate halo masses hosting LABs or spatial clustering
of LABs can be measured, interpreted in the context of the $\Lambda$CDM clustering 
model. 
We also find that the area-halo mass relation:
\begin{equation}
\label{eqn:MA}
{\rm area} = 5.0 \left({M_h\over 10^{12}\msun}\right)^{1.15} {\rm arcsec}^2,
\end{equation}
shown as the solid curve in the right panel of Figure~\ref{fig:halomass}.
Equations (\ref{eqn:ML}) and (\ref{eqn:MA}) lead to the following luminosity-size 
relation
\begin{equation}
\label{eqn:AL}
{\rm area} = 
 5.0 \left({\La\over {10^{42.4}{\rm erg\, s^{-1}}}}\right ) {\rm arcsec}^2,
\end{equation}
which matches the observed one, nothing new in this except as a self-consistency check.

By extrapolating the above relations (2,3) to higher halo mass and using the analytic halo 
mass function \citep[][]{2001Jenkins}, we can obtain the global $\lya$ LF expected
from our model.
In detail, we draw halo masses based on the analytic halo mass function. For each
halo, we compute $\La$ from Equation~(\ref{eqn:ML}). A scatter in $\log\La$
is added following a Gaussian distribution with 1$\sigma$ deviation of 0.28dex
(indicated by the dotted lines in the left panel of Figure~\ref{fig:halomass}). 
Then Equation~(\ref{eqn:AL}) is used to assign the area, and a Gaussian scatter 
of 0.11dex is added to approximately reproduce the scatter seen in the observed
luminosity-size relation. The implied scatter in the area-halo mass relation
is the sum of the above two scatters in quadrature, i.e., about 0.30 dex, which
is indicated by the dotted lines in the right panel of Figure~\ref{fig:halomass}.
Finally, we adopt the same area cut ($>$15 ${\rm arcsec}^2$) used in observations
\citep[][]{2011Matsuda} to define LABs. 

Our computed global $\lya$ LF of LABs is shown as the blue curve in the bottom panel 
(d).
The agreement between our predicted global LF and that from the larger-survey-volume 
observations of 
\citet[][]{2011Matsuda} is striking.
Given still substantial uncertainties involved in our model assumptions,
the precise agreement is not to be overstated.
However, the fact that the relative displacement between LF from our simulated volume and global LF
is in agreement with that between \citet[][]{2004Matsuda} and \citet[][]{2011Matsuda} is
quite encouraging, recalling that we have no freedom to adjust any cosmological parameters. 
This is also indicative of the survey volume 
of \citet[][]{2011Matsuda} having becoming a fair sample of the universe for LABs in question.
The blue dots in top panel (d) show that the predicted luminosity-area
relation is simultaneously in agreement with observations,
now over the entire luminosity and size range,
suggesting that our derived relations in Equations (\ref{eqn:ML}), (\ref{eqn:MA}), 
and (\ref{eqn:AL}) are statistically applicable to LABs of luminosities higher than those probed by the current simulations.

\section{Conclusions and Discussions}

We present a new model, termed star-burst model (SBM),
for the spatially extended (tens to hundreds of kiloparsecs)
luminous ($\La\ga 10^{43}$erg/s) Ly$\alpha$ blobs.
The SBM model is the first model to successfully reproduce
both the global $\lya$ luminosity function and the luminosity-size relation 
of the observed LABs
\citep{2004Matsuda,2011Matsuda}.
In the SBM model $\lya$ emission from both stars and gravitational sources (such as gravitational binding energy released from structure collapse)
is included, although it is found that the nebular $\lya$ emission sourced by those other than stars, 
while significant, is sub-dominant compared to stellar emission.
It is also found that $\lya$ emission originating from sources rather than stars 
is at least as centrally concentrated as that from stars within each galaxy.

Our modeling is based on a high-resolution large-scale cosmological 
hydrodynamic simulation of structure formation,
containing more than $3000$ galaxies with halo mass $M_h>10^{10}\msun$ and
more than $25$ galaxies with $M_h>10^{12}\msun$ at $z=3.1$, all resolved at a resolution of $159$pc or better.
Detailed 3D $\lya$ radiative transfer calculation is applied to sub-volumes centered 
on each of the 40 most massive star-bursting galaxies in the simulation box with ${\rm SFR}=10-400\msun$yr$^{-1}$. 
A self-consistent working model emerges, if proper dust attenuation trend is modeled
in that $\lya$ emission from higher SFR galaxies are more heavily attenuated by dust than lower SFR galaxies, 
which is empirically motivated by observations.
For the results shown, we adopt an effective $\lya$ optical depth $\tau({\rm SFR}) = 0.2 [{\rm SFR}/(\msun{\rm yr^{-1}})]^{0.6}$,
which translates to escape fractions of ($5\%$, $45\%$) for $\lya$ photons at ${\rm SFR}=(100,10) \msun{\rm yr}^{-1}$, respectively.
The dust attenuation model has two parameters, a normalization and a powerlaw index.
The powerlaw index actually follows the slope of the metal column density dependence on SFR in the simulation. 
This thus leaves us with the normalization as the only free parameter.
In practice, changing the powerlaw index does not sensitively change the results, as long as
the normalization is adjusted such that the attenuation at high SFR end ($\sim 100\msun$yr$^{-1}$) 
is approximately the same as the adopted value, making the model rather robust.

Also very encouraging is that the model is in broad agreeement with other observed properties of LABs, 
in addition to the simultaneous reproduction of the observed 
global $\lya$ luminosity function and the luminosity-size relation aforenoted.
Among them, we predict that LABs at high redshift correspond to
proto-clusters containing the most massive galaxies/halos in the universe
and ubiquitous strong infrared emitters, with the most luminous member galaxies mostly 
copious in FIR emission, fully consistent with extant observations \citep[e.g.,][]{2007Geach, 2012Bridge}. 
It seems inevitable that some of the galaxies would contain active galactic nuclei (AGN) at the epoch of peak AGN formation in the universe
\citep[e.g.,][]{2009Geach}. 
While it is straight-forward to include, the results shown do not 
include AGN, partly because, to the zero-th order, we may simply ``absorb" that by adjusting the dust attenuation effect
and partly because observations indicate AGN contribution is subdominant
\citep[e.g.,][]{2009Webb, 2011Colbert}.

The most massive halos in the standard cold dark matter universe
also tend be the most strongly clustered
in the universe, among all types of galaxies,
and we predict that there should be numerous galaxies clustered around LABs \citep[e.g.,][]{2008Uchimoto}. 
\citet{2012Prescott} use high-resolution {\it Hubble Space Telescope} imaging
to resolve galaxies within a giant LAB at $z\sim 2.656$. 
They find many compact, low-luminosity galaxies. 
Their observation becomes incomplete below $\sim 0.1L^*$, 
and with extrapolation there would be about 80 sources above $0.01L^*$ 
within a radius of 7$\arcsec$.
Their LAB has $\La = 10^{44} {\rm erg\, s^{-1}}$ and an isophotal area 
$\sim 140 \,{\rm arcsec}^2$, falling well onto the observed luminosity-size 
relation shown in Figure~\ref{fig:pilot2}. Extrapolating from our model,
the LAB is predicted to reside in a halo with mass of $\sim 10^{13}M_\odot$
(Fig.~\ref{fig:halomass}). The number of faint sources within 7$\arcsec$ above $0.01L^*$ 
from our model would be about 100, in agreement with the observation. 
With the availability of ALMA, observers could make use of its superb capabilities to confirm 
the generic prediction of this model that there should be FIR sources
in each LAB with the most luminous FIR source likely representing the center of the proto-cluster.
In combination with optical and other observations, this will potentially provide extremely useful information on the formation
of galaxies in the most overdense regions of the universe when star formation is most vigorous and clusters
have yet to be assembled.

We highlight here that a potentially very discriminating signature of this model lies in 
the expected, significant polarization strength of the $\lya$ emission at large scales 
($\sim$10−100kpc), which is not expected in some competing models for LABs, such as those 
sourcing primarily gravitational binding energy on large scales due to massive halo formation.
We plan to quantify 
this signal with detailed polarization radiative transfer calculations of $\lya$ photons. 

It is mentioned in passing that our model suggests the trends seen in LABs, in terms of 
the global $\lya$ luminosity function and the luminosity-size relation of the observed LABs,
are continuously extended to less luminous $\lya$ emitters (LAEs).
Consequently, we predict that LAEs, less luminous than LABs,
have smaller sizes compared to those of LABs at a fixed isophotal level
and should also be less strongly clustered than LABs,
forming an extension of the observed LAB luminosity-size relation as well as the LAB luminosity and correlation functions.

Finally, it is reassuring to note that the cosmological simulations themselves
have already been subject to and passed a range of tests concerning a variety of observables of galaxies and the intergalactic medium,
including properties of DLAs at $z=0-4$ \citep[][]{2012Cen},
O~VI absorbers in the circumgalactic and intergalactic medium in the local universe \citep[][]{2012bCen},
global evolution of star formation rate density and cosmic downsizing of galaxies \citep[][]{2011bCen},
galaxy luminosity functions from $z=0$ to $z=3$ \citep[][]{2011bCen, 2011cCen}, 
and properties of galaxy pairs as a function of environment in the low-$z$ universe 
\citep[][]{2012Tonnesen}, among others.

\vskip 1cm

We would like to thank Dr. Yuichi Matsuda for kindly providing 
and allowing us to use new observational data before publication.
Computing resources were in part provided by the NASA High-End Computing 
(HEC) Program through the NASA Advanced
Supercomputing (NAS) Division at Ames Research Center.
R.C. is supported in part by grant NNX11AI23G. Z.Z. is supported in part by NSF
grant AST-1208891. The simulation data are available from the authors upon request.


\end{document}